\documentclass[aps,prc,twocolumn,showpacs,letterpaper]{revtex4}
\usepackage{amssymb}

\usepackage{amsmath,amssymb,mathrsfs}
\usepackage{graphicx}
\usepackage{amsthm}
\usepackage{amscd}
\usepackage{bm}
\usepackage{dsfont}
\usepackage{epsfig}

\def\D{D \hspace{-0.7em}/}

\usepackage{lscape}
\newcount\panelwidth \newcount\panelheight 
\newcount\bxmin \newcount\bymin \newcount\bxmax \newcount\bymax
\newcount\tbxmin \newcount\tbymin
\newcount\tpanelwidth \newcount\tpanelheight \newcount\tpdif
\def\panelsize #1,#2;{\panelwidth=#1 \panelheight=#2}  
\def\setbb #1,#2;#3,#4;#5,#6;{
  \tbxmin=#1 \tbymin=#2    
  \bxmin=#3 \bymin=#4      
  \bxmax=#5 \bymax=#6}     
\def\barepanel #1{%
  \ifnum\panelheight=0 
    \tpdif=\bymax \advance\tpdif by -\bymin
    \multiply \tpdif by \panelwidth
    \tpanelheight=\tpdif
    \tpdif=\bxmax \advance\tpdif by -\bxmin
    \divide \tpanelheight by \tpdif
  \else \tpanelheight=\panelheight \fi
  \ifnum\panelwidth=0 
    \tpdif=\bxmax \advance\tpdif by -\bxmin
    \multiply \tpdif by \panelheight
    \tpanelwidth=\tpdif
    \tpdif=\bymax \advance\tpdif by -\bymin
    \divide \tpanelwidth by \tpdif
  \else \tpanelwidth=\panelwidth \fi
  \epsfig{file=#1,silent=,%
     bbllx=\bxmin bp,bblly=\bymin bp,bburx=\bxmax bp,bbury=\bymax bp,clip=,%
     width=\tpanelwidth mm,height=\tpanelheight mm}}
\def\labelypanel #1{
  \ifnum\panelheight=0 
    \tpdif=\bymax \advance\tpdif by -\bymin
    \multiply \tpdif by \panelwidth
    \tpanelheight=\tpdif
    \tpdif=\bxmax \advance\tpdif by -\bxmin
    \divide \tpanelheight by \tpdif
  \else \tpanelheight=\panelheight \fi
  \ifnum\panelwidth=0 
    \tpdif=\bxmax \advance\tpdif by -\bxmin
    \multiply \tpdif by \panelheight
    \tpanelwidth=\tpdif
    \tpdif=\bymax \advance\tpdif by -\bymin
    \divide \tpanelwidth by \tpdif
  \else \tpanelwidth=\panelwidth \fi
  \tpdif=\bxmax \advance\tpdif by -\tbxmin
  \multiply \tpanelwidth by \tpdif
  \tpdif=\bxmax \advance\tpdif by -\bxmin
  \divide \tpanelwidth by \tpdif
  \epsfig{file=#1,silent=,%
    bbllx=\tbxmin bp,bblly=\bymin bp,bburx=\bxmax bp,bbury=\bymax bp,%
    clip=,width=\tpanelwidth mm,height=\tpanelheight mm}}
\def\labelxpanel #1{%
  \ifnum\panelheight=0 
    \tpdif=\bymax \advance\tpdif by -\bymin
    \multiply \tpdif by \panelwidth
    \tpanelheight=\tpdif
    \tpdif=\bxmax \advance\tpdif by -\bxmin
    \divide \tpanelheight by \tpdif
  \else \tpanelheight=\panelheight \fi
  \ifnum\panelwidth=0 
    \tpdif=\bxmax \advance\tpdif by -\bxmin
    \multiply \tpdif by \panelheight
    \tpanelwidth=\tpdif
    \tpdif=\bymax \advance\tpdif by -\bymin
    \divide \tpanelwidth by \tpdif
  \else \tpanelwidth=\panelwidth \fi
  \tpdif=\bymax \advance\tpdif by -\tbymin
  \multiply \tpanelheight by \tpdif
  \tpdif=\bymax \advance\tpdif by -\bymin
  \divide \tpanelheight by \tpdif
  \epsfig{file=#1,silent=,%
    bbllx=\bxmin bp,bblly=\tbymin bp,bburx=\bxmax bp,bbury=\bymax bp,%
    clip=,width=\tpanelwidth mm,height=\tpanelheight mm}}
\def\labelxypanel #1{%
  \ifnum\panelheight=0 
    \tpdif=\bymax \advance\tpdif by -\bymin
    \multiply \tpdif by \panelwidth
    \tpanelheight=\tpdif
    \tpdif=\bxmax \advance\tpdif by -\bxmin
    \divide \tpanelheight by \tpdif
  \else \tpanelheight=\panelheight \fi
  \ifnum\panelwidth=0 
    \tpdif=\bxmax \advance\tpdif by -\bxmin
    \multiply \tpdif by \panelheight
    \tpanelwidth=\tpdif
    \tpdif=\bymax \advance\tpdif by -\bymin
    \divide \tpanelwidth by \tpdif
  \else \tpanelwidth=\panelwidth \fi
  \tpdif=\bxmax \advance\tpdif by -\tbxmin
  \multiply \tpanelwidth by \tpdif
  \tpdif=\bxmax \advance\tpdif by -\bxmin
  \divide \tpanelwidth by \tpdif 
  \tpdif=\bymax \advance\tpdif by -\tbymin 
  \multiply \tpanelheight by \tpdif
  \tpdif=\bymax \advance\tpdif by -\bymin
  \divide \tpanelheight by \tpdif
  \epsfig{file=#1,silent=,%
    bbllx=\tbxmin bp,bblly=\tbymin bp,bburx=\bxmax bp,bbury=\bymax bp,%
    clip=,width=\tpanelwidth mm,height=\tpanelheight mm}}

\begin{document}
\parskip=0ex 


\title{Neutrino propagation in nuclear medium and neutrinoless double-beta decay}

\author{S. Kovalenko$^{1}$, M. I.\ Krivoruchenko$^{2,3}$ and  F. \v{S}imkovic$^{4,5,6}$}
\affiliation{
$^{1}$Universidad T\'{e}cnica Federico Santa Mar\'ia,
Centro-Cientifico-Tecnol\'{o}gico de Valparaiso\\
Casilla 110-V, Valparaiso, Chile\\
$^{2}$Institute for Theoretical and Experimental Physics$\mathrm{,}$ B.
Cheremushkinskaya 25\\
117218 Moscow, Russia\\
$^{3}$Department of Nano-$\mathrm{,}$ Bio-$\mathrm{,}$ Information and Cognitive Technologies\\
Moscow Institute of Physics and Technology$\mathrm{,}$ 9 Institutskii per.\\
141700 Dolgoprudny$\mathrm{,}$ Russia \\
$^{4}$Bogoliubov Laboratory of Theoretical Physics$\mathrm{,}$ JINR\\
141980 Dubna$\mathrm{,}$ Russia\\
$^{5}$Department of Nuclear Physics and Biophysics\\
Comenius University$\mathrm{,}$ Mlynsk\'a dolina F1\\
SK--842 48 Bratislava$\mathrm{,}$ Slovakia\\
$^{6}$ Institute of Experimental and Applied Physics\\
Czech Technical University in Prague \\ 
128--00 Prague, Czech Republic
}

\begin{abstract}
We discuss a novel effect in neutrinoless double beta ($0\nu\beta\beta$) decay related with the fact that its underlying mechanisms take place 
in the nuclear matter environment. We study the neutrino exchange mechanism and demonstrate the possible impact of nuclear medium via Lepton Number Violating (LNV) 4-fermion interactions of neutrino with quarks from decaying nucleus. The net effect of these interactions is generation of an effective in-medium Majorana neutrino mass matrix. The enhanced rate of the $0\nu\beta\beta$-decay can lead to the apparent incompatibility of observations of the $0\nu\beta\beta$-decay with the value of the neutrino mass determined or restricted by the $\beta $-decay and cosmological data. The effective neutrino masses and mixing are calculated for the complete set of the relevant 4-fermion neutrino-quark operators. Using experimental data on the $0\nu\beta\beta$ decay in combination with the $\beta$-decay and cosmological data we evaluate the characteristic scales of these operators: $\Lambda_{LNV} \geq 2.4$ TeV. 
\end{abstract}

\maketitle

Various mechanisms of neutrinoless double beta ($0\nu\beta\beta$) decay have been considered in the literature (for recent reviews see \cite{VERG12},  \cite{Deppisch:2012nb}). The mechanisms are conventionally constructed 
as Lepton Number Violating (LNV) quark-lepton processes proceeding in vacuum. Then after an appropriate hadronization the presence of the initial and final  nuclei is taken into account as a smearing effect via convolution with the corresponding nuclear wavefunction. On the other hand the nuclear matter may impact an underlying LNV process in a more direct way via the Standard Model (SM) or beyond the SM interactions. If this is relevant, an especially notable effect should be expected from the LNV interactions with the nuclear matter. In the present letter we consider the Majorana neutrino 
exchange mechanism and examine possible impact of nuclear medium via LNV 4-fermion Neutral Current interactions of neutrino with quarks from decaying nucleus. The nuclear matter effect on the $0\nu\beta\beta$-decay rate is calculated in the mean field approach. The mean field associated with the strong interaction is created in nuclei by the scalar and vector quark currents and described effectively in terms of the $\sigma $- and $\omega $-mesons \cite{CHIN74}. Here we consider the scalar mean field associated with the LNV interaction. Then an effective  4-fermion neutrino-quark \mbox{Lagrangian} with the operators of the lowest dimension can written in the form:
\begin{eqnarray}
{\cal L}_{eff} &=&  \frac{1}{\Lambda^{2}_{LNV}}   \sum\limits_{i, j, q}\left(g^{q}_{ij} \overline{\nu_{Li}^{C}}  \nu_{Lj} \cdot  \bar{q}  q +
\mbox{h.c.}\right)+ \label{EffLag} \\
&+& \frac{1}{\Lambda^{3}}   \sum\limits_{i, j, q} h^{q}_{ij} \overline{\nu_{Li}}  i \gamma^{\mu}\overleftrightarrow{\partial}_{\mu} \nu_{Lj} \cdot  \bar{q}  q \, ,
%
\label{EffLag-bis}
\end{eqnarray}
where the fields
$\nu_{Li}$ are the active neutrino left-handed flavor states, $g^{q}_{ij}$ and $h^{q}_{ij}$ are their dimensionless  couplings to the scalar quark currents with $i,j=e,\mu, \tau$ satisfying
$ g^{q}_{ij} = g^{q}_{ji}$ and  $(h^{q}_{ij})^{*} = h^{q}_{ji}$.
The first property follows from the identity \mbox{$\overline{\nu_{Li}^{C}}  \nu_{Lj} = \overline{\nu_{Lj}^{C}}  \nu_{Li}$}, the second one from the hermiticity  of the neutrino operator in the form of kinetic term.
Note that first term in Eq. (\ref{EffLag}) violates lepton number by two units $\Delta L = 2$ while the second one is lepton number conserving  $\Delta L = 0$.  We neglect all the surface terms, which could, in principle, be nontrivial due to the presence of nuclear surface where the gradient of the nuclear matter density is large.  Thus we consider a simplified case of the infinite nuclear radius. 
The scales $\Lambda_{LNV}$ and $\Lambda$ of the $\Delta L = 2$ and $\Delta L = 0$ operators are in general different and are of the order of the masses $M$ of virtual particles inducing these effective operators at tree level.  These particles could be either scalars or vectors (vector leptoquarks) with the masses 
\mbox{$M\gg p_{F} \sim 280$ MeV}, where $p_{F}$ is the Fermi momentum of nucleons in nuclei, which sets the momentum scale of 
$0\nu\beta\beta$-decay. Gauge invariant structure of the operators in Eq. (\ref{EffLag}) is briefly discussed later. 

In the mean field approximation we replace the operator $\bar{q}q$ in Eq. (\ref{EffLag}) with its average value $\langle\bar{q}q\rangle$ over the nuclear medium.  Relying on the MIT 
bag model we have for the light quarks $q=u, d$ an estimate 
$\langle\bar{q}q\rangle \approx \frac{1}{2}\langle q^{\dagger}q \rangle$  \cite{JAFF80}, which is equivalent to
$\langle\bar{q}q\rangle  \approx 0.25 \; \mathrm{fm}^{-3}$  at the saturation.
Thus in the nuclear environment the Lagrangian (\ref{EffLag}) is reduced to 
\begin{eqnarray}\label{EffLag-red}
{\cal L}_{eff} &=&  \frac{\langle\bar{q}q\rangle}{\Lambda^{2}_{LNV}}  \left(\overline{\nu_{Li}^{C}} \ g_{ij}\  \nu_{Lj} + \mbox{h.c.}\right)+\\
\nonumber
&+& \frac{\langle\bar{q}q\rangle}{\Lambda^{3}}  \overline{\nu_{Li}} h_{ij}  i \gamma^{\mu}\overleftrightarrow{\partial}_{\mu} \nu_{Lj} 
 \, ,
\end{eqnarray}
where
 \mbox{$g_{ij} = (g^{u}_{ij} + g^{d}_{ij})/2$} and
\mbox{$h_{ij} =  (h^{u}_{ij} + h^{d}_{ij})/2$}. We assume for simplicity the nuclear medium to be isosinglet.

Let us remind the terms of the electroweak Lagrangian in vacuum relevant to the calculation of the amplitude of 
$0\nu\beta\beta$-decay via Majorana neutrino exchange mechanism. They are
\begin{eqnarray}\label{EW-Lag}
{\cal L}^{vac}_{EW} &=&  \frac{1}{4}    \overline{\nu_{Li}}  i \gamma^{\mu}\overleftrightarrow{\partial}_{\mu} \nu_{Li}  -   
\frac{1}{2} \overline{\nu_{Li}^{C}}\  \hat{M}^{L}_{ij} \  \nu_{Lj} + 
\\
\nonumber
&+& \frac{4 G_{F}\cos\theta_{C}}{\sqrt{2}}\   \overline{l_{Li}}  \gamma^{\mu} \nu_{Lj} \cdot  \overline{u}_{L} \gamma_{\mu} d_{L}  \ \ + \ \ \mbox{h.c.}  \, ,
\end{eqnarray}
where $M^{L}_{ij} =  M^{L}_{ji} $ is a Majorana mass matrix symmetric by the same reason as $h^{q}_{ij}$ matrix in Eqs. (\ref{EffLag}) and (\ref{EffLag-bis}). It can be diagonalized by a unitary transformation
$
\nu_{i} = U^{L}_{ij}\nu'_{j}.
$
In the basis where the charged lepton mass matrix is diagonal the unitary matrix $U^{L}$ coincides with the PMNS mixing matrix. 
Thus in the vacuum we have
\begin{eqnarray}\label{EW-Lag-mass-basis}
{\cal L}^{vac}_{EW} &=&  \frac{1}{4}    \overline{\nu^{\prime}_{Li}}  i \gamma^{\mu}\overleftrightarrow{\partial}_{\mu} \nu^{\prime}_{Li}  -   
\frac{1}{2} m_{i}\ \overline{\nu^{\prime \ C}_{Li}}\  \nu^{\prime}_{Li} + 
\\
\nonumber
&+& \frac{4 G_{F}\cos\theta_{C}}{\sqrt{2}} \  \overline{l_{Li}}  \gamma^{\mu}\  U^{L}_{ij}\  \nu^{\prime}_{Lj} \cdot  \bar{u}_{L}\gamma_{\mu} d_{L}  \ \ + \ \ \mbox{h.c}  \, .
\end{eqnarray}
Here, $m_i$ (i=1,2,3) is the neutrino mass in the vacuum.

According to the conventional parameterization $U^{L} = V^{L} D$, where $V^{L}$ is a matrix depending on the three mixing angles and one Dirac phase,  
\mbox{$D = Diag\{1, \exp (i \alpha_{21}/2), \exp (i \alpha_{31}/2)\}$} is the diagonal matrix of Majorana phases, which are chosen so that 
\mbox{$m^{*}_{i} = m_{i} \geq 0$} and  the entry $V^{L}_{e3} = \sin^{2}\theta_{13}$ has no the Dirac phase.

As seen from Eq. (\ref{EffLag-red}) the neutrino interactions with the nuclear matter affect both the mass and kinetic terms of the vacuum Lagrangian
(\ref{EW-Lag}), (\ref{EW-Lag-mass-basis}) so that the in-medium Lagrangian written in the vacuum mass eigenstate basis takes the form:
\begin{eqnarray}\label{EW-Lag-mass-inmedium}
{\cal L}^{\mathrm{med}}_{EW} =  
\frac{1}{4}    \overline{\nu^{\prime}_{Li}} \ \widehat{\cal K}_{ij} \  i \gamma^{\mu}\overleftrightarrow{\partial}_{\mu} \nu^{\prime}_{Lj}  -   
\frac{1}{2} \overline{\nu^{\prime \ C}_{Li}}\  \widehat{\cal M}_{ij} \  \nu^{\prime}_{Lj} + 
\nonumber\\
+ \frac{4 G_{F}\cos\theta_{C}}{\sqrt{2}} \  \overline{l_{Li}}  \gamma^{\mu}\  U^{L}_{ij}\  \nu^{\prime}_{Lj} \cdot  \bar{u}_{L} \gamma_{\mu} d_{L}  \ \ + \ \ \mbox{h.c.}  \, ,
\end{eqnarray}
where 
\begin{eqnarray}\label{def-1}
\widehat{\cal K}_{ij} = \delta_{ij} + 4 \frac{\langle \bar{q} q\rangle}{\Lambda^{3}} \hat{h}_{ij},~
\widehat{\cal M}_{ij} = m_{i} \delta_{ij} -2 \frac{\langle \bar{q} q\rangle}{\Lambda_{LNV}^{2}} \hat{g}_{ij},
\end{eqnarray}
with  $\hat{h} = U^{L \dagger} h\  U^{L}$,  $\hat{g} = \left(U^{L}\right)^{T} g \ U^{L}$. We thus have $
\widehat{\cal K}^{\dagger} = \widehat{\cal K}$, and $ \widehat{\cal M}^{T} = \widehat{\cal M}$.

First we bring the neutrino kinetic term in the Lagrangian (\ref{EW-Lag-mass-inmedium}) to the canonical form. Towards this end 
we diagonalize it by a unitary transformation
$
\nu^{\prime}_{i} = V_{ij}\nu^{\prime\prime}_{j}$, $V^{\dagger} \widehat{\cal K} V = Diag\left\{ \lambda_{k}  \right\}$$ \equiv \Omega
$,
where $\lambda^{*}_{k} = \lambda_{k} \geq 0 $.   The positiveness of these eigenvalues is maintained as long as 
$
4 \langle \bar{q} q \rangle \hat{h} \leq \Lambda^{3}
$,
which is implied in our analysis. With this condition a field rescaling 
$\nu^{\prime\prime}_{i} \rightarrow  \lambda^{-1/2}_{i}\nu^{\prime\prime}_{i}$
%
%
allows us to arrive at the canonical kinetic term
\begin{eqnarray}\label{EW-Lag-mass-inmedium-1}
\nonumber
{\cal L}^{\mathrm{med}}_{EW} &=&  
\frac{1}{4}    \overline{\nu^{\prime\prime}_{Li}}  \  i \gamma^{\mu}\overleftrightarrow{\partial}_{\mu} \nu^{\prime\prime}_{Li}  -   \\
\nonumber
&-& \frac{1}{2} \overline{\nu^{\prime\prime \ C}_{Li}}\ \lambda^{-1/2}_{i} V_{ji} \widehat{\cal M}_{jk} V_{kn} \lambda^{{-1/2}}_{j}  
\nu^{\prime\prime}_{Ln} + 
\\
\nonumber
&+& \frac{4 G_{F}\cos\theta_{C}}{\sqrt{2}} \  \overline{l_{Li}}  \gamma^{\mu}\  U^{L}_{ij}V_{jk}\lambda^{-1/2}_{k} \nu^{\prime\prime}_{Lk} 
\cdot  \bar{u}_{L} \gamma_{\mu} d_{L}  \\
&+& \ \ \mbox{h.c.}  
\end{eqnarray}
Then  we diagonalize the effective Majorana mass term by a unitary transformation $\nu^{\prime\prime}_{i} = W^{L}_{ij} \tilde\nu_{j}$,
\begin{equation}\label{Diag-fin}
\left(W^{L}\right)^{T} \left(\Omega^{-1/2} V^{T} \widehat{\cal M} V \Omega^{-1/2}\right)W^{L} = 
Diag\left\{\bar\mu_{i}\right\}
\end{equation}
where $\bar\mu_{i} = \mu_{i}\exp(-i\phi_{i})$ with $|\bar\mu_{i}| = \mu_{i}$. These phases can be absorbed by the neutrino fields
\mbox{$\tilde\nu_{Li} \rightarrow \exp(i\phi_{i}/2) \tilde\nu_{Li}$}.
Only two of these phases are physical.  One of $\phi_{1,2,3}$ can be erased  by an overall phase rotation of the charged lepton fields:
\mbox{$l_{L i} \rightarrow l_{L i} \exp(-i\phi_{1}/2)$},
where we conventionally selected the phase $\phi_{1}$ to be eliminated. After all that we finally arrive at the neutrino Lagrangian in the nuclear matter:
\begin{eqnarray}\label{EW-Lag-matter-diagonal}
{\cal L}^{\mathrm{med}}_{EW} &=&  \frac{1}{4}    \overline{\tilde\nu_{Li}}  i \gamma^{\mu}\overleftrightarrow{\partial}_{\mu} \tilde\nu_{Li}  -   
\frac{1}{2} \mu_{i}\ \overline{\tilde\nu^{C}_{Li}}\  \tilde\nu_{Li} + 
\\
\nonumber
&+& \frac{4 G_{F}\cos\theta_{C}}{\sqrt{2}} \  \overline{l_{Li}}  \gamma^{\mu}\  U^{\mathrm{eff}}_{ij}\  \tilde\nu_{Lj} \cdot  \bar{u}_{L} \gamma_{\mu} d_{L}  \ \ + \ \ \mbox{h.c}  \, .
\end{eqnarray}
in terms of an effective mass eigenstate neutrino fields $\tilde\nu_{Li}$ in the nuclear environment related to the in-vacuum fields $\nu_{i}$ from 
Eq. (\ref{EW-Lag}) as $\nu_{L i} = U^{\mathrm{eff}}_{ij} \tilde\nu_{j}$
with $U^{\mathrm{eff}} = U^{L} V \Omega^{-1/2} W^{L}{\cal P}$,
where ${\cal P} = Diag\left\{1, \exp (i\phi_{21}/2),  \exp (i\phi_{31}/2)]\right\}$ is the diagonal matter generated Majorana phase matrix,
with $\phi_{21} = \phi_{2} - \phi_{1}$, $\phi_{31} = \phi_{3} - \phi_{1}$.
Note that the neutrino mixing matrix in medium  $U^{\mathrm{eff}}$ is not unitary, contrasting to unitarity of the neutrino mixing matrix 
$U^{L}$ in vacuum. 

The amplitude of $0\nu\beta\beta$-decay for the Majorana neutrino exchange in nuclear medium is proportional to
the quantity
%
\begin{equation}
m_{\beta \beta }=\sum_{i}(U_{ei}^{\mathrm{eff}})^{2}\mu _{i},  \label{meff}
\end{equation}
which should be compared with the corresponding quantity without nuclear matter effects
\begin{eqnarray}\label{meff-vac}
m^{vac}_{\beta \beta }=\sum_{i}(U_{ei}^{L})^{2}m_{i}.
\end{eqnarray}

\vspace{-9mm}
\begin{center}
\begin{figure}[htbp]            
  \panelsize 92,0;               
  \setbb 0,0; 0,0; 580,470;    
  \labelypanel{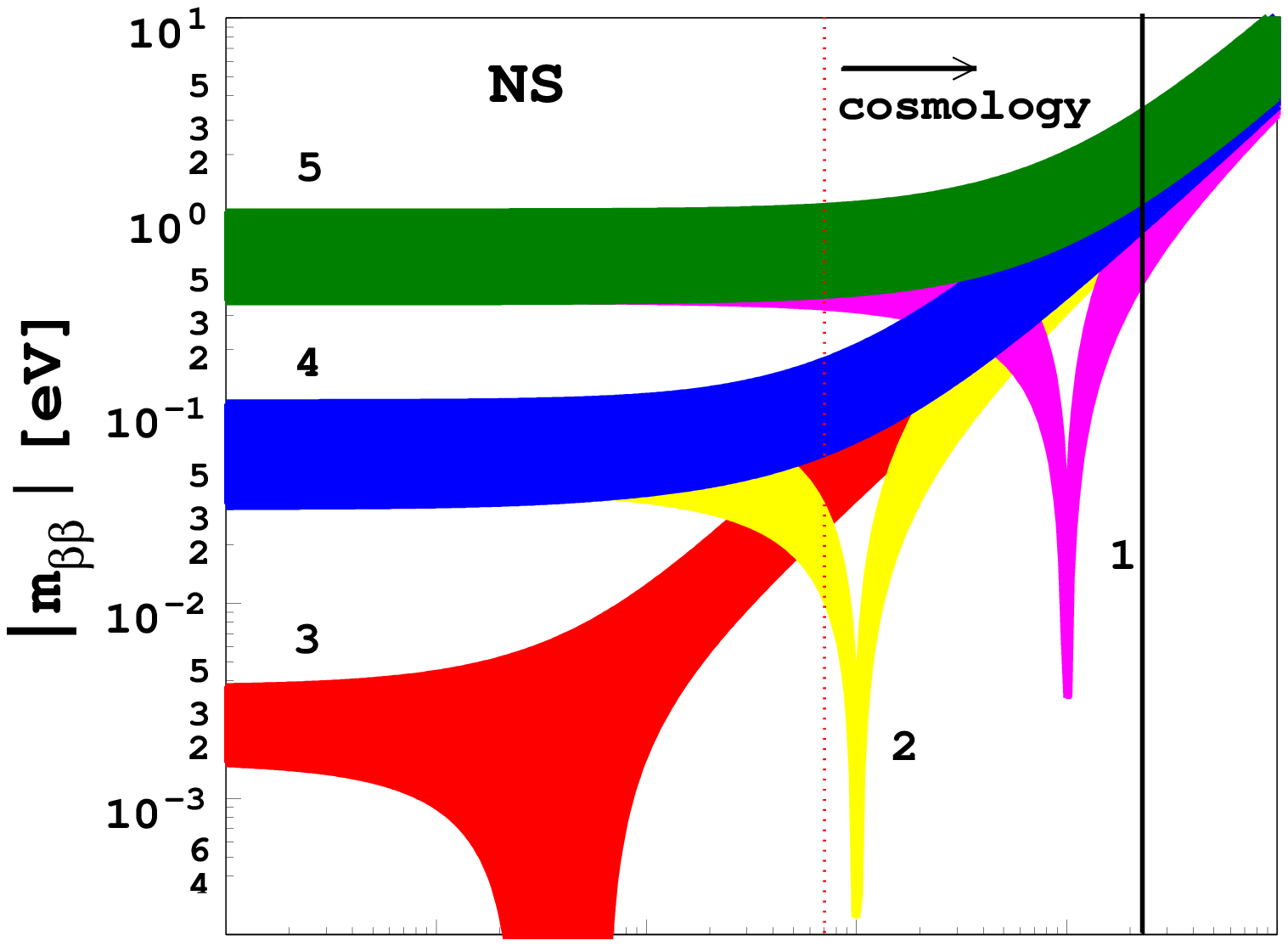} \\   
  \vspace{-25.5mm}
  \labelxypanel{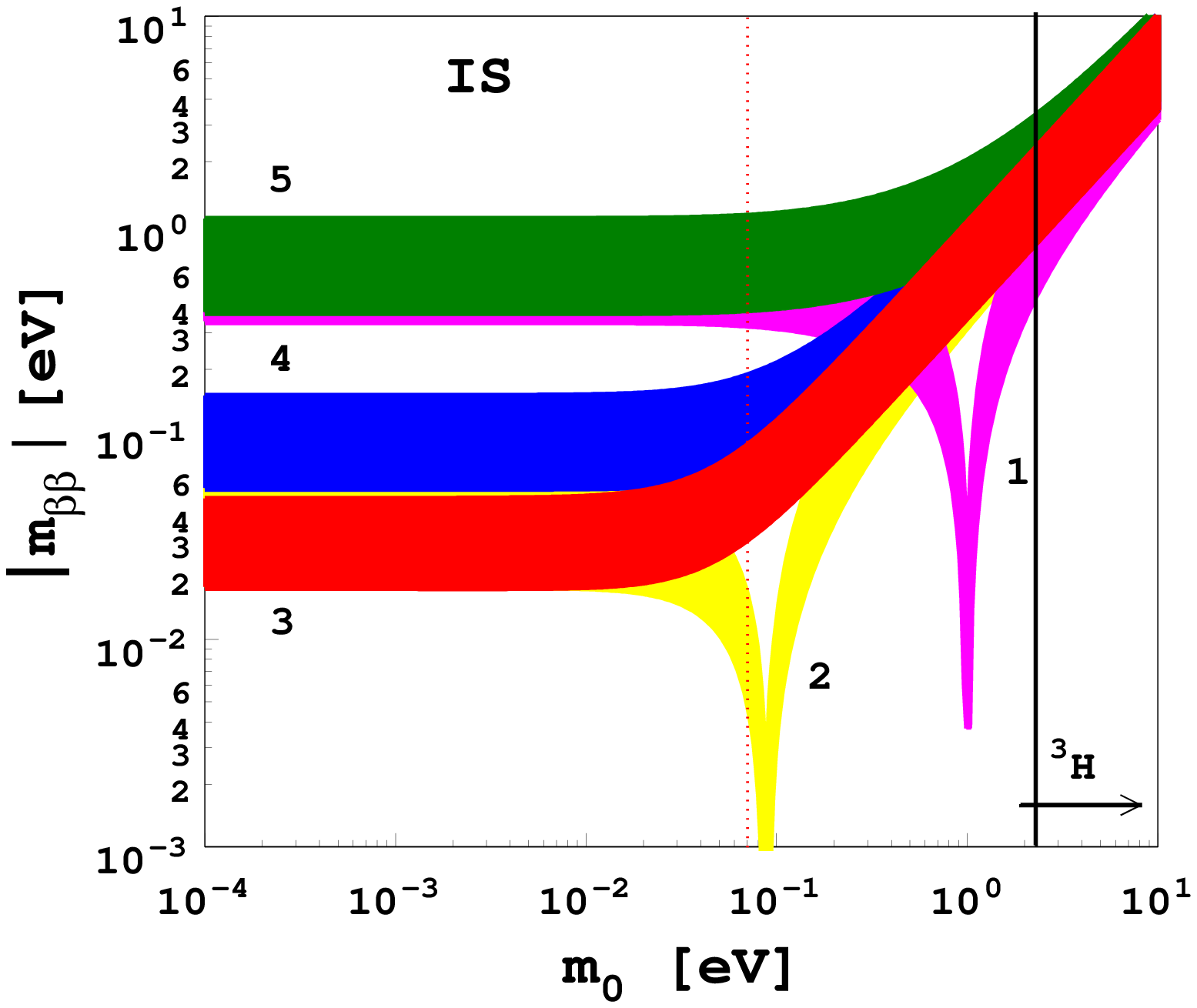}
  \vspace{-9mm}
\caption{(color online)
The bands 1, 2, 3, 4 and 5 show admissible values of $|m_{\beta \beta }|$ 
and $m_0$ for $h=0$ and $\langle \bar{q} q \rangle g =-1$, $-0.1$, $0$, $0.1$, and $1$
eV, respectively. 
The upper and lower panels correspond to the normal (NS) 
and the inverted (IS) neutrino spectrums.
%
The CP phases spread in the interval [0,2$\pi$]. 
Regions to the right from the vertical  solid and dotted lines are excluded by the tritium $\beta$-decay \cite{TRIT11} 
and by the cosmological data \cite{PLAN13,THOM10}.
}
\label{fig-2}
\end{figure}                    
\end{center}

The experimental searches for $0\nu\beta\beta$ decay provide information on the in-medium effective parameter $m_{\beta \beta} $ from Eq. (\ref{meff}).
%
For various choices of nuclear matrix elements the currently most stringent limit on this parameter derived by 
EXO-200 and KamLAND-Zen experiments with $^{136}$Xe \cite{136XE12}
and by GERDA experiment with  $^{76}$Ge \cite{GERDA} is in the range 
$
| m_{\beta \beta }| \leq 0.2 - 0.4~ \mathrm{eV}.
$
Discussion of the next-generation experiments aimed at improving 
the $0\nu\beta\beta$ limits can be found in Ref. \cite{VERG12}.

The information on the in-vacuum neutrino masses and mixing is provided by neutrino oscillation experiments (for a review see \cite{PDG12}).
The quantities measured in these experiments are the neutrino mass squared differences \mbox{$\Delta m^{2}_{ij} = m^{2}_{i} - m^{2}_{j}$}
and mixing angles $\theta_{12}, \theta_{23}$ and $\theta_{13}$.
If the overall mass scale is fixed, e.g., by the mass of the
lightest neutrino, $m_{0} \equiv \min (m_{i})$, all the other masses are
determined. 
Two types of the neutrino mass spectra are possible: the normal one with $ m_{1}<m_{2}<m_{3}$ (NS)  
and the inverted one with $ m_{3}<m_{1}<m_{2}$ (IS).

The overall neutrino mass scale in vacuum can be constrained by tritium beta decay measurements and cosmological data.

The presently best experimental limit on the neutrino parameter $m_{\beta}$ observable in tritium beta decay is \cite{TRIT11}:
$m_{\beta}^2 = \sum_{i} |U^{L}_{ei}|^{2} m_{i}^{2} \leq$ (2.2 eV$)^2$ at 95\% C.L.
The KATRIN experiment  is expected to improve this limit by a factor of 10 in the near future \cite{Drexlin:2013lha}. 

Recently, the Planck collaboration \cite{PLAN13} reported new limits on the sum of the neutrino masses:
$
\sum_{i} m_i \le 0.23 - 1.08 ~\textrm{eV}
$, 
derived from the measurements of the temperature 
of the cosmic microwave background and lensing-potential power spectra.
The lowermost bound implies \mbox{$m_0 \le 0.07~ \mathrm{eV}$}. 
An upper limit of $0.28 - 0.47$ eV for the sum of neutrino masses was reported in Ref. \cite{THOM10}.

From the constraints of Refs. \cite{136XE12,GERDA} and
\cite{TRIT11,PLAN13,THOM10} we derive limitations on the 4-fermion effective neutrino-quark interactions 
introduced in Eq. (\ref{EffLag}). 
We consider a simplified case for the scalar couplings in Eqs. (\ref{EffLag})-(\ref{EffLag-red}) such that
$4 \hat{h}_{ij}{\Lambda^{-3}}=\delta _{ij}h$, 
$2 \hat{g}_{ij}{\Lambda^{-2}_{LNV}} = \delta _{ij}g$,
with $h, g$ being real numbers, where  $\hat{h}, \hat{g}$ are defined after Eq. (\ref{def-1}).
Then  we have
$V_{ij} = \delta _{ij}$, 
$W^{L}_{ij} = \delta _{ij}$,  
$\Omega_{ij} = \delta_{ij} \lambda$, 
$\lambda = 1 + \langle \bar{q} q\rangle h $, 
$\mu_{i} = \lambda^{-1} \left| m_{i} - \langle \bar{q} q\rangle g\right|$.
The effective Majorana mass (\ref{meff}) in this case is
\begin{eqnarray}
m_{\beta \beta } =  \sum^n_{i=1} \left(V_{ei}^{L }\right)^{2}\xi_{i} 
\frac{| m_{i} - \langle\bar{q} q \rangle g |}{\left(1 - \langle\bar{q} q\rangle h\right)^{2}}.
\label{mmed}
\end{eqnarray}
Here $V^{L}_{ij}$ is the PMNS mixing matrix in vacuum without Majorana phases.  
The Majorana phase factor is
\mbox{$\xi_{i} = \{1, \exp(i\alpha_{1}), \exp(i\alpha_{2}) \}$} with \mbox{$\alpha_{1} = (\alpha_{21} + \phi_{21})/2$}, 
$\alpha_{2} = (\alpha_{31} + \phi_{31})/2$, 
where $\alpha_{ij}$ are the Majorana phases in vacuum defined together with the matrix 
$V^{L}$ after  Eq. (\ref{EW-Lag-mass-basis}).
Within the simplified scheme, the quantity $m_{\beta\beta}$  in nuclear medium in comparison with the one in vacuum depends on the two new unknown parameters:  $ h, g$. 
In our numerical estimations we assume that only one of them is different from zero at a time.
The unknown phases in Eq. (\ref{mmed}) are varied in the interval $[0, 2\pi]$. The vacuum mixing angles and the neutrino mass squared differences are taken from 
Ref. \cite{PDG12}.
We illustrate our results in Fig.~\ref{fig-2}.
The shaded areas display allowed values of $|m_{\beta \beta}|$ and
$m_{0}$ for a set of sample values of $g$ with $h=0$.
For both NS and IS these results, being combined with the cosmological and tritium $\beta$-decay limits, suggest for the LNV scale
\begin{eqnarray}\label{Lim-from-g1}
\Lambda_{LNV} \geq 2.4 \, \mbox{TeV}\, \mbox{(Planck)}, \ \  1.1 \, \mbox{TeV}\, \mbox{(Tritium)}
\end{eqnarray} 
%
With the future KATRIN data the limit 1.1 TeV in Eq. (\ref{Lim-from-g1}) will be pushed up to \mbox{$\sim$ 2 TeV}.   For convenience we also give our limits 
in terms of a dimensionless parameter $\varepsilon_{ij}$  defined as 
$\varepsilon_{ij} G_{F}/\sqrt{2} = g_{ij}/\Lambda^{2}_{LNV}$ and characterizing the relative strength of the 4-fermion LNV operators in (\ref{EffLag}) 
with respect to the Fermi constant $G_{F}$. From (\ref{Lim-from-g1}) we have 
$\varepsilon_{ij} \leq 0.02$ (Planck), $0.1$ (Tritium).


The effect of a nonzero value of the coupling constant $h \neq 0$ is particularly simple. 
Its variation results in shifting the plots in Fig.~\ref{fig-2} along the vertical axis. For the case $g= 0$, corresponding to the domains 3 
in Fig.~\ref{fig-2}, the limit 
$m_{\beta\beta} \leq 0.2$ eV  implies very weak constraint $\Lambda \geq 0.2$ GeV on the scale $\Lambda$ of the  
Lepton Number conserving operator in Eq. (\ref{EffLag}).

Let us briefly comment on the gauge invariant origin of the operators in Eq. (\ref{EffLag}).  The lepton number conserving operator with the derivative stems after the Electroweak Symmetry Breaking (EWSB) from the SM gauge invariant operators of the type:
\begin{eqnarray}\label{Deriv-GI}
%
\frac{1}{M^{4}} \left[ z^{d}\bar{L} i \D\ L \cdot \bar{Q} d_{R} \cdot H + z^{u}\bar{L} i \D\   L \cdot \bar{Q} u_{R}  \cdot  \epsilon H^{\dagger}\right],
\end{eqnarray}
where  $L$, $Q$ and $H$ are the lepton, quark and Higgs $SU_{2L}$ doublets. The gauge invariant contractions of their components are implied and involve 
the SM gauge covariant derivative $\D = \gamma_{\mu} D^{\mu}$. 
%
The LNV operators in Eq. (\ref{EffLag}) may have various origins. Some examples are:
\begin{eqnarray}\label{LNV-GI}
%
&&\frac{1}{M_{LNV}^{3}} \left[ \kappa_{1}\overline{L_{\alpha}^{C}} L_{\beta} \cdot\bar{Q}_{\alpha} u_{R}\cdot  \epsilon_{\beta\gamma}H_{\gamma}\right. +
\\   
\nonumber
&&\hspace{12mm}\left. + \kappa_{2}\overline{L^{C}_{\alpha}} \gamma_{\mu} u_{R} \cdot  \bar{Q}_{\alpha} \gamma^{\mu}  L_{\beta} \cdot \epsilon_{\beta\gamma}H_{\gamma}+...\right].
\end{eqnarray} 
Here the subscript Greek letters denote components of the $SU_{2L}$ doublets. 
A complete list of the corresponding operators and their possible ultraviolet completions will be presented elsewhere. 
In Eqs. \eqref{Deriv-GI}, \eqref{LNV-GI} we introduced common scales $M, M_{LNV}$ of the operators and their dimensionless couplings $z^{u,d}, \kappa_{i}$, which are, in general, non-diagonal matrices in the flavor space. 
After the EWSB due to $\langle H^{0}\rangle = v$ these operators engender the corresponding operators in Eqs. \eqref{EffLag},  \eqref{EffLag-bis} with the scales 
\begin{eqnarray}\label{Scales-Rel}
\frac{z^{q} v}{M^{4}} = \frac{h^{q}}{\Lambda^{3}},  \ \ \ \frac{\kappa v}{M_{LNV}^{3}}  = \frac{g}{\Lambda_{LNV}^{2}}.
\end{eqnarray}
 

As seen from Eq. (\ref{LNV-GI}), due to the gauge invariance the terms with the scalar  quark currents $\bar{q} q $  in Eq. (\ref{EffLag}), appearing after the EWSB, have to be accompanied with the pseudoscalar ones $\bar{q} \gamma_{5} q$ having the same couplings so that:
\begin{eqnarray*}
\frac{1}{\Lambda_{LNV}^{2}} \overline{\nu_{L}^{C}} \nu_{L} \left[(g^{u} \bar{u} u  + g^{d} \bar{d} d) +
(g^{u} \bar{u} \gamma_{5}u  - g^{d} \bar{d} \gamma_{5} d)  \right].
\end{eqnarray*}
Therefore,  the scale $\Lambda_{LNV}$  can also be evaluated from $BR(\pi^{0}\rightarrow \nu \nu)\leq 2.7\times 10^{-7}$ \cite{PDG12}. Assuming $g_{ij} = 1$ as in 
Eq. (\ref{Lim-from-g1}) we have $\Lambda_{LNV}\geq 560$ GeV, which is less stringent than those in Eq. (\ref{Lim-from-g1}).

Note that the scale  (\ref{Lim-from-g1}) of the operators in Eq. (\ref{EffLag-bis}) suggests underlying renormalizable mechanisms with heavy intermediate particles 
with masses at the TeV scale which is within the reach of the experiments at the LHC. As shown in Ref. \cite{Helo:2013ika} these experiments 
have great potential in distinguishing the underlying mechanisms and setting limits on the scales of the effective operators. 


Non-standard interactions affect neutrino propagation in matter. Thus, one may expect additional constraints on the energy scale of these interactions from astrophysical implications. The vector 4-fermion interactions $\bar\nu\gamma\nu\cdot \bar{q}\gamma q$ are intensively discussed in the literature (for a review see Ref. \cite{Ohlsson:2012kf}). Their contribution to the in-medium neutrino Hamiltonian is independent of the neutrino energy in contrast to the scalar-type interactions whose effect reduces to renormalization of the neutrino mass matrix suppressed by the neutrino energy. Neutrino oscillations in matter are therefore much less sensitive to the interaction of Eqs. (\ref{EffLag}) and (\ref{EffLag-bis}). On the other hand, our constraints (\ref{Lim-from-g1}) are comparable to the most stringent ones derived so far for the non-standard interactions of the vector type.


Note that the Majorana neutrino mass $m_{\nu}$ in vacuum and the LNV operators in Eq. (\ref{EffLag})  should originate from the same underlying LNV physics at 
the energy scales above $\Lambda_{LNV}$.  However mechanisms generating these two effective Lagrangian terms may be very different.  In this context
it is instructive to estimate the significance of the direct contribution $\delta m_{\nu}$ of the LNV operators in Eq. (\ref{EffLag}) to the Majorana neutrino mass.
This contribution is given by the quark bubble attached to the neutrino line as it follows from the contraction of the quark fields in Eq. (\ref{EffLag}).
The result is 
$
\delta m_{\nu} \sim {g^{q}} / 
(4\pi \Lambda_{LNV})^{2}  m_{q}^{3} \log({\Lambda_{LNV}}/{m_{q}})
$
where $m_{q}$ is the light quark $q=u, d$ mass in the loop. The usual $\overline{MS}$ renormalization scheme is applied to obtain the finite result. This relies on the assumption that 
the complete underlying theory  is renormalizable. For $\Lambda_{LNV} \sim 2.4$ TeV, $m_{d} \sim 5$ MeV and $g^{q}=1$ we find $\delta m_{\nu} \sim 10^{-6}$ eV, which is very small 
and could represent only a subdominant contribution to neutrino mass. There must be another mechanism of the neutrino mass generation compatible with the neutrino oscillation data. 

In the future, the gradually improving cosmological and single $\beta$-decay neutrino mass limits may come into conflict with the possible evidence of $0\nu\beta\beta$ decay. If so, the new physics would be mandatory. 
In particular it can be represented by the new effective TeV scale neutrino-quark interactions (\ref{EffLag}), (\ref{EffLag-bis}) enhanced in $0\nu\beta\beta$ decay by the nuclear mean field. 
If the dominant mechanism of $0\nu\beta\beta$ decay is Majorana neutrino exchange, the scenario presented here will provide the most direct explanation for the above mentioned possible
incompatibility between the experiments. 

In conclusion, we revisited the Majorana neutrino exchange mechanism of $0\nu\beta\beta$-decay  in the presence of 
non-standard LNV interactions of neutrino with nuclear matter of decaying nucleus. 
These interactions were parametrized with the effective Lepton Number Violating and Lepton Number conserving  4-fermion neutrino-quark  operators of  
the lowest dimension. In terms of these operators we calculated the in-medium Majorana neutrino mass, mixing matrix and the parameter
$m_{\beta\beta}$ driving the $0\nu\beta\beta$ decay within the neutrino exchange mechanism.
%
Combining experimental limits on this parameter with the cosmological and tritium beta decay constraints on the neutrino overall mass scale we extracted 
a stringent limit on the scale of the LNV interactions of neutrino with the  quark scalar current. 
In the similar way the nuclear matter may affect 
other underlying mechanisms of $0\nu\beta\beta$ decay. 

\vspace{2mm}

M.I.K. and F.\v{S}. acknowledge kind hospitality at the Universidad T\'{e}cnica Federico Santa Mar\'ia, Valpara\'iso. 
This work was supported in part by RFBR grant No. 13-02-01442, the VEGA Grant agency of the Slovak 
Republic under the contract No. 1/0876/12, and by the Ministry of Education, Youth and Sports of the Czech Republic under contract LM2011027, 
by Fondecyt (Chile) under Grant No. 1100582. 
\\[5mm]

\end{document}